\documentclass[graybox,natbib,nosecnum]{svmult}
\bibpunct{(}{)}{;}{a}{}{,} 

\pdfoutput=1   

\usepackage{mathptmx}       
\usepackage{helvet}         
\usepackage{courier}        
\usepackage{type1cm}        
\usepackage{subfig}
\usepackage{makeidx}         
\usepackage{graphicx}        
\usepackage{multicol}        
\usepackage[bottom]{footmisc}
\usepackage[normalem]{ulem}	
\usepackage{hyperref}  
\usepackage{floatrow}
\usepackage{soul}   

\makeindex             

\begin{document}

\title*{Internal Structure of Giant and Icy Planets: Importance of Heavy Elements and Mixing}
\author{Ravit Helled and Tristan Guillot}
\institute{Ravit Helled \at Institute for Computational Sciences, University of Zurich, Winterthurerstr.~190
CH 8057 Zurich, Switzerland. \email{rhelled@physik.uzh.ch}
\and Tristan Guillot \at Observatoire de la Cote dAzur, Bd de
lÕObservatoire, CS 34229, 06304 Nice Cedex 4, France. \email{tristan.guillot@oca.eu}}
%
%
\maketitle

\abstract{\\
In this chapter we summarize current knowledge of the internal structure of giant planets. We concentrate on the importance of heavy elements and their role in determining the planetary composition and internal structure, in planet formation, and during the planetary  long-term evolution.  
We briefly discuss how internal structure models are derived, present the possible structures of the outer planets in the Solar System, and summarise giant planet formation and evolution. 
Finally, we introduce giant exoplanets and discuss how they can be used to better understand giant planets as a class of planetary objects. }

\section{Introduction}
Characterisation of the outer planets in the Solar System has been one of the major objectives in planetary science since decades. 
Throughout the years significant progress has been made, both in theory and observations. We now have a much better understanding of the behaviour of hydrogen and other elements at high pressures and temperatures, and the physical processes that govern the planetary structure. The various spacecrafts that have visited (and are currently visiting) the outer planets in the Solar System, Jupiter, Saturn, Uranus, and Neptune, provide us with constraints on the gravitational fields, rotation periods, and atmospheric compositions of the planets that can be used by structure models. 
In parallel, the discovery of giant planets around other stars (giant exoplanets) provides an opportunity to study the diversity in giant planet composition, which can be used to better understand giant planet formation. 
\par

Despite the great progress in planetary modelling in the last few decades there are still several open questions regarding the nature of Jupiter, Saturn, Uranus and Neptune. 
Many review chapters have been written recently on giant planet interiors (e.g., Fortney \& Nettelmann, 2010, Guillot \& Gautier, 2014; Baraffe et al., 2014, Militzer et al., 2016) and this chapter aims to be somewhat complementary to those.  Our chapter is organised as follows. 
First, we discuss the interiors of the Solar System's gas giant planets (Jupiter and Saturn) and icy planets (Uranus and Neptune). Second, we discuss the standard formation mechanism of giant planets and how it is linked to their composition. 
Finally, we provide an outlook on the compositions of giant exoplanets. 

\section{Giant Planet Structure}
\subsection{Making an interior model}
Information on the interiors of the giant planets in the Solar System is typically derived from theoretical structure models which are designed to fit the observed physical data of the planets, such as their gravitational fields, masses, internal rotations, and radii. 
The physical properties used by interiors models of the outer planets are listed in Table 1. 
The planetary interior is modelled by using the following structure equations  which include the mass conservation, hydrostatic, thermodynamic, and energy conservation equations:
\begin{equation}
{\partial\over\partial m}{4\pi\over3}r^3={1\over\rho},
\end{equation}
\begin{equation}
{\partial p\over\partial m}=-{Gm\over4\pi r^4} + {\omega^2\over 6\pi r} + {G M\over 4\pi \ R^3 r}\varphi_\omega,
\end{equation}
\begin{equation}
{\partial T\over\partial m}=\nabla_T{\partial p\over\partial m},
\end{equation}
\begin{equation}
{\partial u\over\partial t}+p{\partial\over\partial t}{1\over\rho}
=q-{\partial L\over\partial m},
\end{equation}
where $P$ is the pressure, $\rho$ is the density, $m$ is the mass, $r$ is the radius and $G$ is the gravitational constant. The temperature gradient $\nabla_T$ depends on the process by which the internal heat is transported. 
The last equations is the only equation that is time ($t$) dependent and is used for modelling the planetary evolution. $u$ is the internal energy, $q$ is an energy source that is typically assumed to be zero for planets, and $L$ is the intrinsic luminosity. 
\par 

In order to account for rotation, the hydrostatic equation (Eq.~2) includes additional terms which depend on $\omega$, the spin rate, $M$ the total mass of the planet, $R$ the total radius and $\phi_\omega$ is a function of the radius, internal density and spin rate  (see Guillot 2005). For a non-spinning planet, $\phi_\omega=\omega=0$. For a spinning planet, this equation is valid in the limit of a barotropic fluid and a solid-body rotation. The radius is then considered as a mean volumetric radius. In that case, we can obtain constraints on the internal density distribution by measuring the departure of the planet's gravity field from sphericity. These are expressed in the form of the gravitational moments, even functions of the radius, $r$, and the colatitude, $\theta$ (see e.g., Guillot 2005, Hubbard 2013):
\begin{equation}\label{jn}
J_{2\ell}=-\frac{1}{Ma^{2\ell}}\int r'^{2\ell}P_{2\ell}(\cos\theta')\rho(r',\theta')d^3r'
\end{equation}
where $a$ is the equatorial radius, and $P_n$ is the $n$th-order Legendre polynomial.  Interior models are constructed to fit the mass (essentially $J_0$) and as many of the $J_{2\ell}$'s as have been measured.  Although each higher order $J$ gives additional information on $\rho(r)$. 
The density distribution correspond to a hydrostatic configuration when the contribution of dynamical effects (e.g., winds) on the gravitational moments are not included.   
\par

Unfortunately, there is no unique solution for the internal structure of a planet. The inferred structure depends on the model assumptions and the equations of state (EoSs) used by the modeller.  
The main uncertainties in structure models are linked to the following assumptions/setups: (i) number of layers (ii) the composition and distribution of heavy elements (iii) heat transport mechanism, and (iv) rotation period and the dynamical contribution of winds (e.g., differential rotation). 
\par

Since the gas giant planets (Jupiter and Saturn) consist of mainly hydrogen and helium,  their modelling relies on the EoS of hydrogen, helium, and their mixture. 
The major uncertainty concerning the EoS of hydrogen is in the region of 0.5-10 Mbar, where hydrogen undergoes a transition from a molecular phase to a metallic phase. 
The EoS of helium in the relevant pressure region is simpler since helium ionization requires larger pressures and a phase transition is not expected to occur.
The difficulty with calculating the EoS of helium, however, is due to the separation of helium droplets from the hydrogen-helium mixture (e.g,  Fortney \& Hubbard, 2003; Stevenson \& Salpeter, 1977a,b).  
The EoS for the heavier elements (metals, rocks, ices) have generally received somewhat less attention than those for hydrogen and helium. 
Despite the difficulty, there have been substantial advances in high-pressure experiments and ab initio calculations of EoSs of hyrogen and helium and of  of heavier materials, as well as on the miscibility properties, for water, ammonia, rock, and iron.
Detailed description on EoSs and interior modeling can be found in Saumon \& Guillot (2004), Baraffe et al.~(2014), Fortney \& Nettelmann (2010), Militzer et al.~(2016), Miguel et al.~(2016), Fortney et al.~(2016). 

\begin{table}
\caption{Basic Characteristics of the Outer Planets taken from Guillot \& Gautier (2014) and NASA website: http://ssd.jpl.nasa.gov/?gravity\_fields\_op. Jupiter's gravitational field is taken from Folkner et al., 2017.}
\label{tab:1}   
\begin{tabular}{p{4cm}p{2.4cm}p{2.4cm}p{2.4cm}p{2.4cm}}
\hline\noalign{\smallskip}
 Physical Property & Jupiter& Saturn & Uranus & Neptune  \\
\noalign{\smallskip}\svhline\noalign{\smallskip}
Distance to Sun (au) & 5.204 & 9.582 & 19.201 & 30.047 \\
Mass (10$^{24}$ kg)& 1898.13$\pm$0.19 &  568.319$\pm$0.057 & 86.8103$\pm$0.0087 & 102.410$\pm$0.010\\
Mean Radius (km) & 69911$\pm$6& 58232$\pm$6 & 25362$\pm$7 & 24622$\pm$16 \\
Mean Density (g/cm$^{3}$) & 1.3262$\pm$0.0004 & 0.6871$\pm$0.0002  & 1.270$\pm$0.001 & 1.638$\pm$0.004 \\
$J_2 \times 10^6$ & 14696.51$\pm$0.272 & 16290.71$\pm$0.27  & 3510.68$\pm$0.70 & 3408.43$\pm$4.50\\
$J_4 \times 10^6$ & -586.62$\pm$0.36  & -935.83$\pm$2.77  & -34.17$\pm$1.30 & -33.40$\pm$2.90 \\
$J_6 \times 10^6$ & 34.24$\pm$0.24 & 86.14$\pm$9.64  & ----  & ---- \\
Effective Temperature (K) & 124.4$\pm$0.3 & 95.0$\pm$0.4 & 59.1$\pm$0.3 & 59.3$\pm$0.8\\
1-bar Temperature (K) & 165$\pm$5 & 135$\pm$5  & 76$\pm$2 & 72$\pm$2\\
\noalign{\smallskip}\hline\noalign{\smallskip}
\end{tabular}
\end{table}

\subsection{Jupiter and Saturn}
Typically, the interiors of Jupiter and Saturn are modelled assuming the existence of a distinct heavy-element core which is surrounded by an envelope divided into an inner helium and heavy element rich layer and an outer envelope which is helium poor and less enriched with heavy elements 
(e.g., Guillot, 1999, Saumon \& Guillot, 2004; Nettelmann et al., 2008; 2012).  
The existence of a core is linked to the traditional (and somewhat outdated) view of planet formation in the core accretion scenario, as we discuss below, and the  
division of the envelope into two is based on the idea that at high pressures not only does hydrogen change from the molecular to the metallic phase, but also to the immiscibility of helium in hydrogen (Stevenson \& SalPeter 1977a,b, Fortney \& Hubbard, 2003). 
Recent calculations of the phase diagram of a hydrogen-helium mixture confirm the immiscibility of helium in hydrogen (e.g., Lorenzen et al.,2009; 2011, Morales et al., 2009; 2013). Figure 1 shows the phase diagram for the hydrogen-helium mixture for a helium
mole concentration of 8\% (see Guillot \& Gautier 2014 for details). 
Indeed, the atmospheres of both Jupiter and Saturn are observed to be depleted in helium compared to a proto-solar ratio (von Zahn et al. 1998, Conrath \& Gautier 2000), and helium rain is the most common (although not the only) explanation for Saturn's high thermal emission (see Fortney \& Nettelmann, 2010 and references therein).
The location in which helium rain occurs and its timescale are important to determine the distribution of helium and heavy elements in the interiors of Jupiter and Saturn (e.g., Stevenson \& Salpeter 1977a,b). 
\par

\begin{figure}
\floatbox[{\capbeside\thisfloatsetup{capbesideposition={left,top},capbesidewidth=4cm}}]{figure}[\FBwidth]
{ \caption{\small Phase diagram for a hydrogen-helium mixture. The orange region shows the region of  the H-He separation as derived by Lorenzen et al.~(2011). 
The red curve shows the critical temperature for that separation according to Morales et al. (2013a). Numerical and experimental results by Schouten et al.~(1991) and Loubeyre et al.~(1991) are also presented. 
The back curves show the isentropes of Jupiter (plain) and Saturn (dashed), respectively. The figure is taken from Guillot \& Gautier, 2014.}}
{\includegraphics[scale=.21]{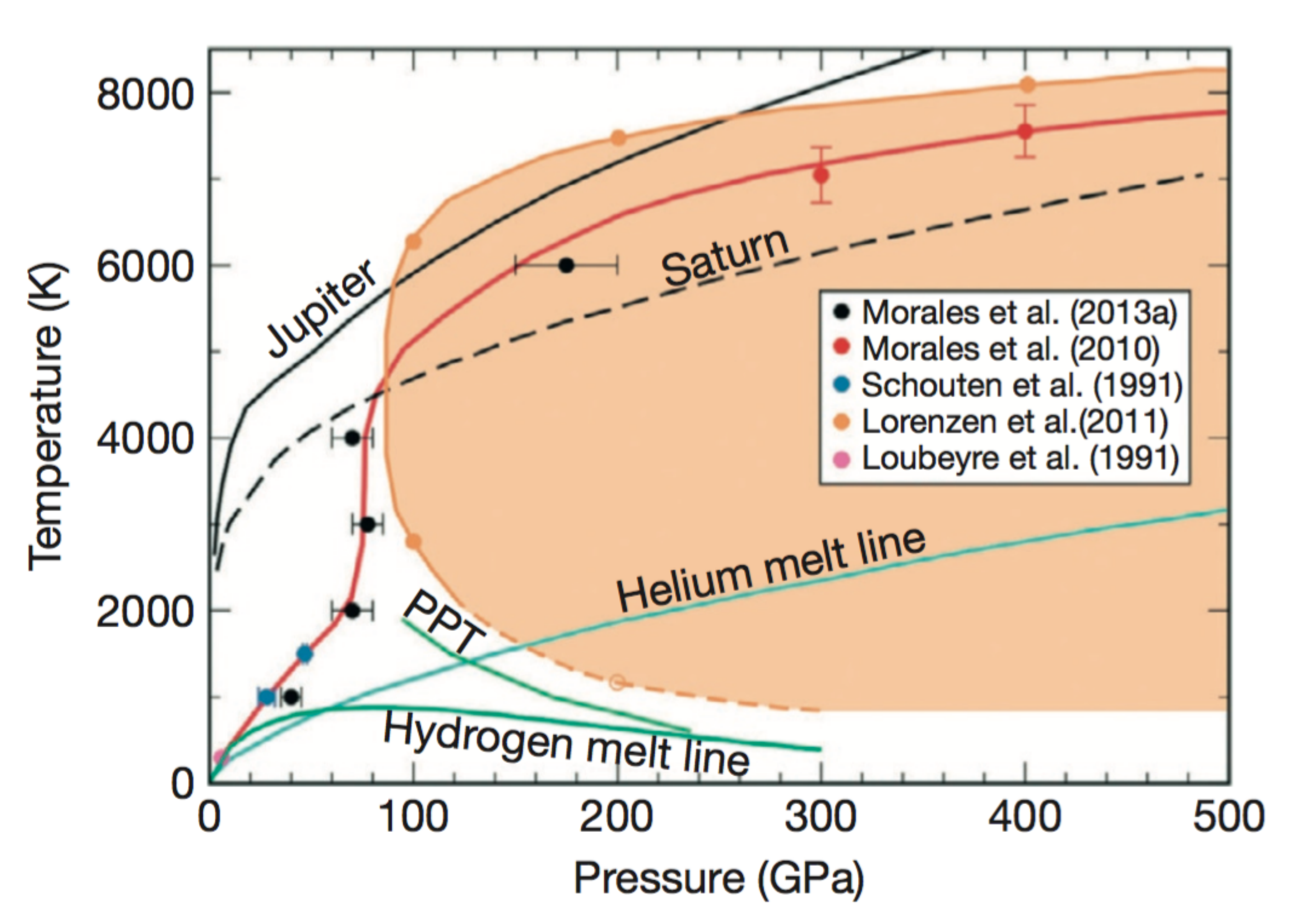}}
\end{figure}

 For Jupiter, standard 3-layer models typically infer a core
mass smaller than $\sim$10 M$_{\oplus}$ (Earth mass). The global enrichment
in heavy elements is uncertain, and the total heavy element mass is estimated to
be between 10 and 40 M$_{\oplus}$ (Saumon \& Guillot, 2004; Nettelmann et al., 2015).  
Alternative models with a different EoS for hydrogen (Militzer et al., 2008; Hubbard \& Militzer, 2016)  
imply the existence of a relatively massive core ($\sim$16 M$_{\oplus}$) and
only a very small enrichment (if at all) in heavy elements in the gaseous envelope.   
Recently, Miguel et al. (2016) investigated the sensitivity of the derived internal structure of Jupiter to the estimates of its gravitational moments and the accuracy of the used EoSs. They suggest that the differences in the inferred structure 
are linked to differences in the internal energy and entropy calculation. This in return leads to differences in the thermal profiles and therefore to different estimations in the core and heavy-element masses.
Overall, it seems that preferable solutions are ones with cores ($\sim$10 M$_{\oplus}$) and a discontinuity of the heavy-element enrichment in the envelope, with the inner helium-rich envelope consists 
of a more heavy elements than the outer, helium-poor envelope. 
Recently, new estimates for Jupiter's gravitational field were determined by the Juno spacecraft (Bolton et al., 2017). 
Interior models of Jupiter that fit the data suggest that another feasible solution for Jupiter is the existence of a diluted core (Wahl et al., 2017). In this case, Jupiter's core is no longer viewed as a pure heavy-element central region with a density discontinuity at the core-envelope-boundary, but as a diluted core which is more extended region, and can also consist of lighter elements.  This model resembles the primordial structure derived by formation models (see below), 
providing a potential link between giant planet formation models, and structure models of the planets at present day.    
\par

The internal structure of Saturn is also uncertain - although its derived structure is less sensitive to the hydrogen EoS (e.g., Saumon \& Guillot, 2004), it is dependent on the hydrogen-helium phase diagram which is not fully constrained.  
Additional complication arises from the uncertainty in Saturn's rotation period and shape (e.g., Fortney et al., 2017).  
Overall, structure models suggest that Saturn is more enriched in heavy elements compared to Jupiter, also having a larger core. 
The total heavy-element mass in Saturn is estimated to be $\sim$16 - 30 M$_{\oplus}$ with a core mass between zero and 20 M$_{\oplus}$ (e.g., Saumon \& Guillot, 2004; Nettelmann et al., 2012; Helled \& Guillot, 2013).

\subsection{Uranus and Neptune}
Uranus and Neptune are the outermost planets in the Solar System. 
Unlike Jupiter and Saturn, their gaseous envelopes are relatively small fractions of their total masses. 
The available constraints on interior models of Uranus and Neptune are limited. The gravitational harmonics of these planets are known only up to fourth degree ($J_2, J_4$), and the planetary shapes and rotation periods are not well determined (e.g., Helled et al. 2010). 
\par

Although Uranus and Neptune have similar masses and radii, they appear to be quite different internally.  The measured low heat flux of Uranus implies that either it has lost its heat or there is a mechanism that reduces the efficient of cooling. 
In addition, Uranus radius is larger than Neptune's but its mass is smaller.
This means that Neptune is denser than Uranus by 30\%. The origin of this dichotomy is unknown, and could be a result of  giant impacts that affected the internal structure of these planets (e.g., Podolak \& Helled, 2012). 
\par

Three main approaches have been used for modelling Uranus and Neptune. The first assumes that the planets consist of three layers: a core made of ``rocks'' (silicates, iron), an ``icy'' shell (H$_2$O, CH$_4$, etc.), and a gaseous envelope (composed of molecular hydrogen and helium with some heavier components). 
This approach uses physical EoSs of the assumed materials to derive a density profile that best fits the measured gravitational coefficients, similarly to the standard models of Jupiter and Saturn (e.g., Nettelmann et al., 2013). 
A second approach makes no a priori assumptions regarding planetary structure and composition. The radial density profiles of Uranus and Neptune that fit their measured gravitational fields are derived using Monte Carlo searches (e.g., Marley et al. 1995, Podolak et al. 2000). 
A third one uses a continuous radial density and pressure profiles that fit the mass, radius, and gravitational moments of Uranus and Neptune, and then use this density profile to investigate the possible composition of the planets by using theoretical EoSs (e.g., Helled et al, 2011). 
While there are variations in the derived composition of Uranus and Neptune using the different approaches several results seem to be robust: all models find that the outer envelopes of the planets are highly enriched with heavy elements, and that the heavy element concentration increases towards the planetary centre.  
\par

Recent 3-layer models suggest that Uranus and Neptune contain a minimum of $\sim$2 M$_{\oplus}$  and about 3 M$_{\oplus}$ of hydrogen and helium, respectively. 
When considering that the planetary interior has distinct layers of different composition (3-layer model), the ice-to-rock ration is found to be high in both planets. The inferred global ice-to-rock-ratio is estimated to be between 19 and 36 in Uranus, while Neptune has a wide range of solutions from 3.6 to 14. 
Random models of Uranus and Neptune suggest both planets consists of small cores and enriched outer envelopes, and that both planets require a density jump at a radius of about 0.6 to 0.7 of the total radius to fit the gravity data (see Marley et al. 1995, Podolak et al. 2000). 
On the other hand, the empirical models of Helled et al.~(2011) suggest that both planets can have a gradual structure in which there is a gradual increase of the heavier material toward the centre. They also found that the innermost regions of both Uranus and Neptune cannot be fit to the empirical density distribution with pure ice/rock, but by $\sim$82\% of SiO$_2$ and $\sim$ 90\% of H$_2$O by mass for both Uranus and Neptune. The overall metallicity of the planets was found to be 0.75-0.92 and 0.76-0.9 for Uranus and Neptune, respectively. 
In addition, they emphasise the fact that the planetary interiors could be depleted in ices, and still fit the measured gravitational field, suggesting that these planets are not necessarily "icy". 
Figure 2 shows the density profiles of the outer planets in the Solar System for standard 3-layer models. 
\par

\begin{figure}%
 \includegraphics[scale=.35]{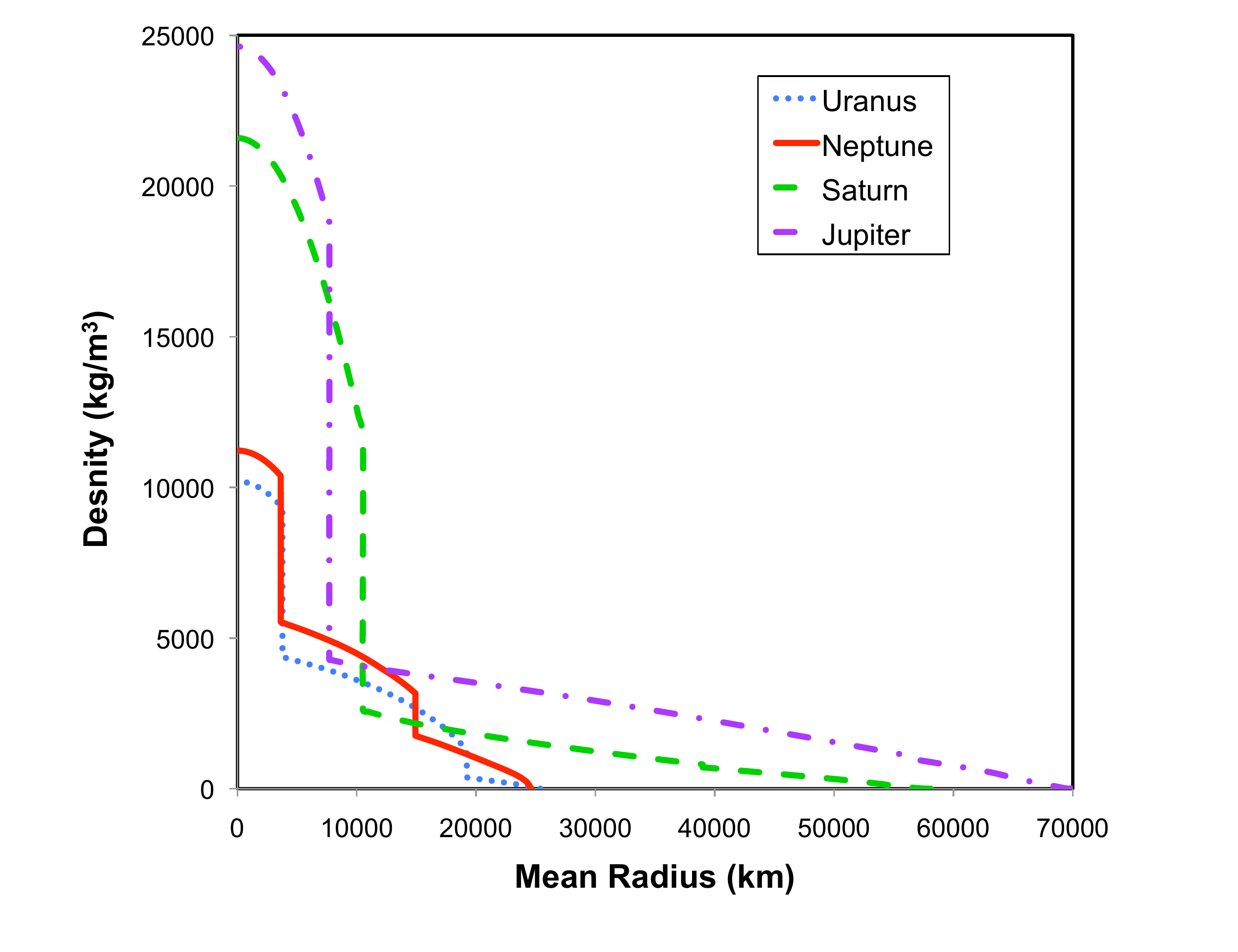}
 \vspace{-0.4cm}
    \caption{
    {\small
     Representative density profiles of Jupiter, Saturn, Uranus and Neptune. Shown is the density as a function of the planetary mean radius. 
The data are taken from Miguel et al., 2016, Helled \& Guillot, 2013, and Nettelmann et al., 2013, respectively.
    }}%
    \label{fig:example}%
\end{figure}

\subsection{Non-Adiabatic Interiors}
We now realise that in some cases, and perhaps in most cases, a fully adiabatic model for the giant planets is too simplistic. The fact that Uranus has a much smaller internal luminosity than Neptune has long ago been attributed to the presence of a molecular weight gradients in the deep interior (Podolak, Hubbard \& Stevenson 1991). The inhibition of convection in the presence of helium rain has also been shown to be a likely possibility (Stevenson \& Salpeter 1977b). Recently, f-mode oscillations of Saturn were discovered through the observation of its rings by the Cassini spacecraft. The analysis of the splitting of these oscillation modes led Fuller (2014) to propose that Saturn's deep interior must be stably stratified: This is at present the only way to explain the unexpected splittings, through interactions between f-modes propagating in the convective envelope and g-modes propagating in the stable region of the deep interior.

A non-adiabatic structure may arise because of a primordial compositional gradient due to the formation process itself, because of the erosion of a central core or because of immiscibility effects (for example of helium in metallic hydrogen). Composition gradients can inhibit convection and affect the heat transport in giant planets. If they are weak and the luminosity is large, they will be overwhelmed by overturning convection which will then ensure a rapid mixing and homogeneization. Otherwise, they can either lead to layered convection, a less efficient type of convection, or inhibit convection and lead to heat transport by conduction or radiation. 

When in the presence of a homogenous composition, the convection criterion is given by the {\it Schwarzschild criterion}, 
$\nabla_{ad}>\nabla$, 
where $\nabla\equiv d\ln T /d\ln P$, and $\nabla_{ad}$ is the adiabatic gradient.
In case of an inhomogeneous environment, one has to take into account the effect of the composition gradient on the stability criterion. Considering a mixture of elements with mass fractions $(X_{1},X_{2,}...,X_{n})$, the composition gradient is given by (e.g., Vazan et al., 2015):
$\nabla_{X}\equiv\frac{\partial\ln T(p,\rho,X)}{\partial X_{j}}\cdot\frac{dX_{j}}{d\ln P}$. 
In this case the convection criterion is given by the {\it Ledoux criterion}, 
$\nabla-\nabla_{ad}-\nabla_{X}<0.$ 

Layered convection is convective mixing that can occur in regions that are stable according to the {\it Ledoux criterion}, but unstable according to the {\it Swcharzschild criterion}, if the entropy and chemical stratifications have opposing contributions to the dynamical stability. In that case, diffusive convection can take place (e.g, Rosenblum et al., 2011; Wood et al., 2013; Mirouh, et al., 2012), leading to slow mixing and a more efficient heat transfer. Layered-convection can occur in two forms: {\it fingering convection} or {\it double-diffusive convection}. In the first, the entropy is stably stratified ($\nabla - \nabla_{ad} < 0$), but the composition gradient is unstably stratified ($\nabla_{X} < 0$); while in the second, oscillatory double-diffusive convection (ODD), entropy is unstably stratified 
($\nabla - \nabla_{ad} > 0$), but chemical composition is stably stratified ($\nabla_{X} > 0$); it is related to semi-convection, but can occur even when the opacity is independent of composition.
\par

 \begin{figure}
 \begin{center}
 \vspace{-2cm}
\includegraphics[scale=.5]{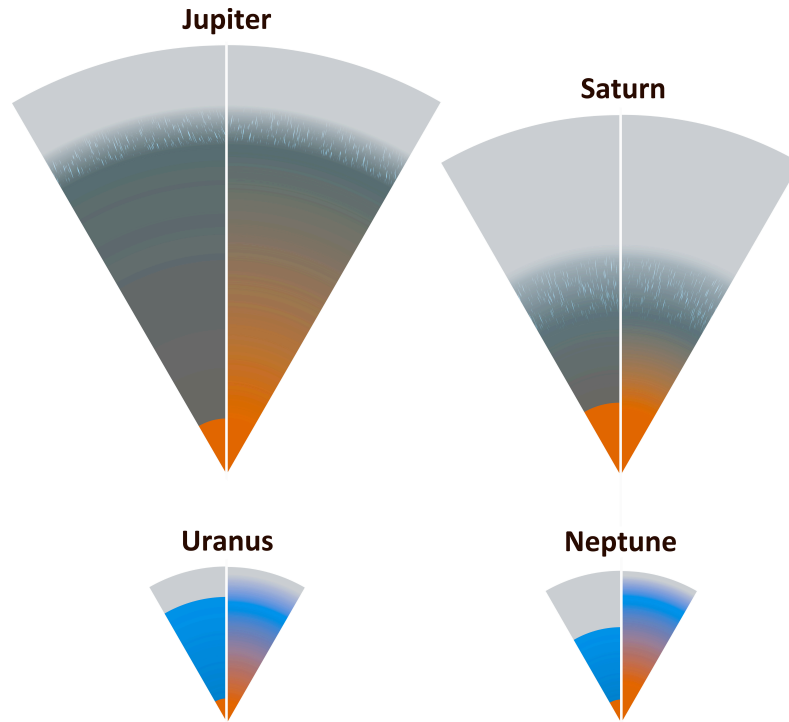}\\
 \vspace{-2.3cm} 
\caption{{\small Sketches of the internal structures of Jupiter, Saturn, Uranus and Neptune as inferred from structure models. For each planet we show two possible structures: one consisting of distinct layers and one with a gradual distribution of heavy elements.}}
\label{fig:1}       
\end{center}
\end{figure}

A pioneering study on double diffusive convection in planetary interiors was presented by Leconte \& Chabrier (2012; 2013) where Jupiter's and Saturn's interiors were modeled assuming the presence of double-diffusive convection caused by a heavy-element gradient in their gaseous envelopes. 
These models investigated the effect on the internal heat transport efficiency and the internal structure. 
In this scenario, the planetary interiors can be much hotter and the planets can accommodate larger amounts (a few tens of $M_{\oplus}$) of heavy elements. 
However, these models can be considered as extreme because they assumed a compositional gradient to be present throughout the envelope. Evolution models matching Jupiter's present constraints show that is almost impossible to avoid overturning convection homogeneizing a large fraction of the envelope (Vazan et al. 2016). This therefore strongly limits the extent and consequences of layered convection. At the same time, the results of Vazan et al. (2016) show that both Jupiter and Saturn can be non-adiabatic and still fit observations.

Evolution models with layered convection in the helium-rain region of Jupiter have recently been calculated (Nettelmann et al. 2015, Mankovich et al. 2016). In these models, the molecular envelope cools over time, but the deep interior can actually heat up because of the loss of specific entropy due to helium settling (see also Stevenson \& Salpeter 1977b). While it is not yet clear that layered convection does really occur in the helium mixing region (it depends on the thermodynamic behavior of the hydrogen-helium mixture in the presence of a phase separation), these models show that non-adiabaticity is certainly an important aspect of the evolution of cool gaseous planets such as Jupiter and Saturn.   

Recently, Nettelmann et al. (2016) modelled Uranus accounting for a boundary layer and modelled the planetary evolution. They find that the existence of such a boundary layer can explain the the low luminosity of the planet. 
The thermal boundary leads to a hotter interior, suggesting that the deep interior could have a large fraction of rocks. 
\par

Investigations of non-adiabatic structure of the outer planets are still ongoing, and we expect major progress in this direction in the upcoming years. 
Overall, each of the  outer planet in the Solar System can be modelled by using a more standard layered interior or alternatively, by a structure in which there is  gradual change in composition. 
Figure 3 shows sketches of the possible internal structures of the outer planets in the Solar System accounting for these two possibilities.

\section{Giant Planet Formation and Heavy Elements Distribution}

\subsection{The core accretion model}
The standard model for giant planet formation is known as {\it core accretion} (see Helled et al., 2014 for a review). 
In this model, the formation of a gaseous planet begins with the buildup of a heavy-element core due to the growth and accretion of solids which can be in the form of planetesimals 
(e.g., Pollack et al., 1996; Alibert et al., 2005) or pebbles (e.g., Lambrechts et al., 2014, Levison et al. 2016) and continues with gas accretion.             
The planetary formation history can be divided into three main phases. The first phase, {\it Phase-1}, is dominated by core heavy-element accretion. 
A small core accretes 
planetesimals/pebbles until it has 
obtained most 
of the heavy-element mass $M_\mathrm{Z}$ within its gravitational reach.  
The gas mass, $M_\mathrm{gas}$ (H-He), also grows, but it
remains only a very small fraction of $M_\mathrm{Z}$.
During the second phase, {\it Phase-2}, a gaseous envelope is being accreted slowly, 
$\dot{M_\mathrm{Z}}$ 
decreases considerably, and $\dot{M_\mathrm{gas}}$ increases slowly until it 
exceeds the heavy-element accretion rate.
As the envelope's mass  increases, the expansion of the zone of
gravitational influence allows further accretion of planetesimals. 

\begin{figure}
\floatbox[{\capbeside\thisfloatsetup{capbesideposition={left,top},capbesidewidth=4cm}}]{figure}[\FBwidth]
{ \caption{{\small An example of the growth of a protoplanet in the core accretion model.
The planet mass is shown vs.~time up to {\it Phase-3} (crossover).  
 \textit{Dotted purple line}: the actual core mass. 
\textit{Dashed black line}: gaseous (hydrogen and helium only) mass. \textit{Solid red line}: heavy-element mass.  \textit{Solid blue line}: total planetary mass.
This simulation corresponds to planet formation at 5.2 AU with a local solid-surface density of 10 g cm$^{-2}$.}}\label{fig:test}}
{\includegraphics[scale=.32]{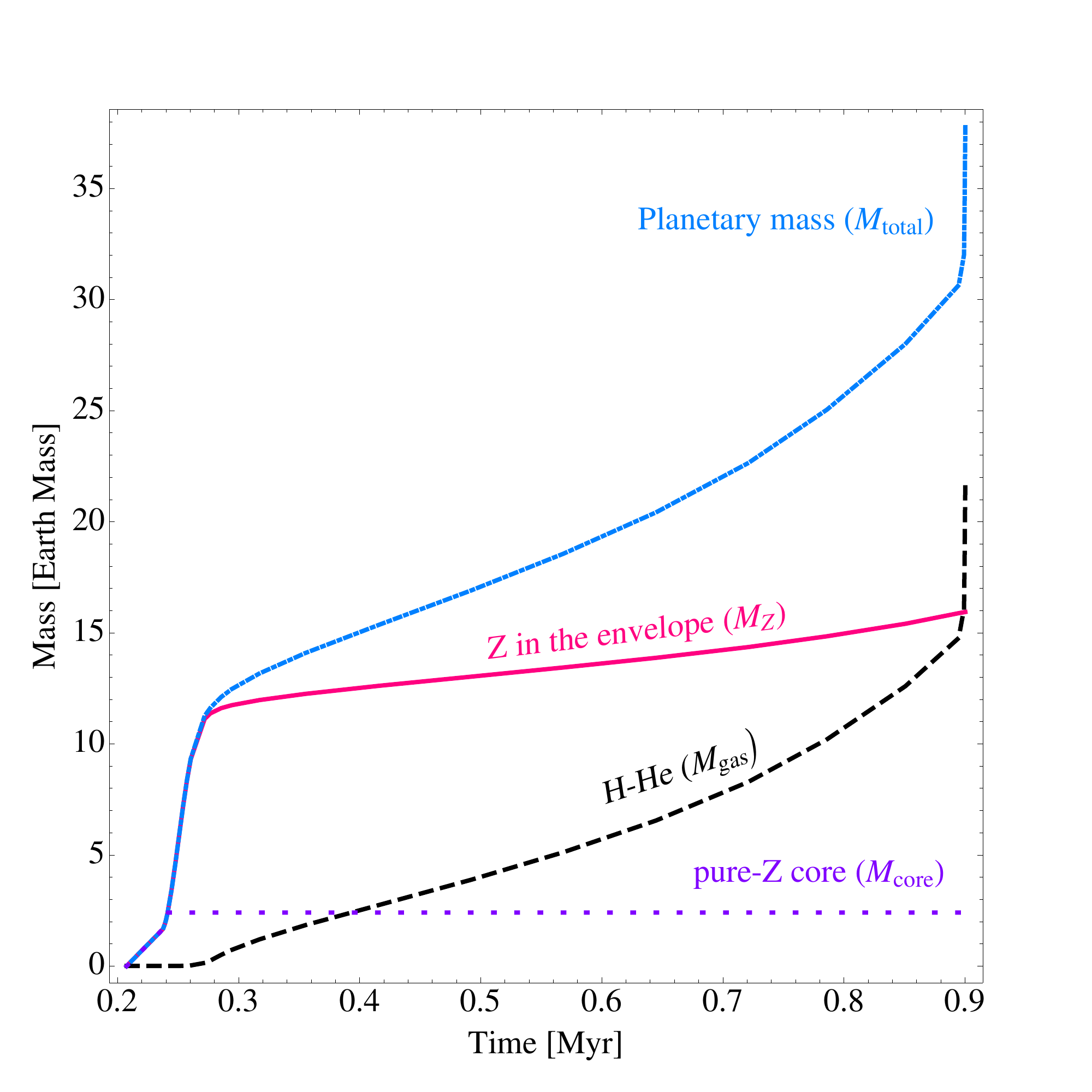}}
\end{figure}

{\it Phase-3} correspond to the runaway gas accretion phase. 
When  $M_\mathrm{Z} \sim M_\mathrm{gas}$, known as crossover, the gas accretion rate increases
considerably, nearly at free fall.  
The heavy-element accretion rate during this phase is poorly known but is typically assumed to be small (Helled \& Lunine, 2014). 
The gas accretion is terminated by either disk dissipation or gap opening, and the planet gains its final mass (assuming no mass loss or late accretion occur).   

\subsection{Core Growth and Mixing}
The envelope of the forming giant planets is typically considered to consist mainly of hydrogen and helium.  
If the accreted heavy-elements reach the center (core) without depositing mass in the envelope, the planetary envelope has a sub-stellar composition due to the depletion in heavies, and in this case $M_\mathrm{env} \sim M_\mathrm{gas}$. 
However, if planetesimals/pebbles suffer a strong
mass ablation as they path through the gaseous envelope they can lead to a substantial enrichment with heavy elements, typically resulting in a metal-rich proto-atmosphere. 
In this case, the core mass $M_\mathrm{core}$ and the heavy element mass $M_\mathrm{Z}$ can differ.
The determination of the planetesimal mass ablation depends on the characteristics of the accreted heavy elements such as their composition, size,  
and mechanical strength. Formation models typically find that when $M_\mathrm{core} \sim 2\rm\, M_{\oplus}$ the solids tend to remain in the atmosphere 
(Iaroslavitz \& Podolak, 2007; Lozovsky et al., 2017). 
The enrichment of the planetary envelope (envelope pollution) has a strong influence on the planetary growth; 
it can 
strongly reduce the critical mass of the planet for triggering rapid gas accretion, i.e., in reaching {\it Phase-3} (e.g., Hori \& Ikoma, 2011; Venturini et al., 2016). 
\par

It is not clear at this point whether the last phase of accretion, in which most of the planetary mass is gained, is that of heavy-element poor gas or whether heavy elements manage to be accreted very efficiently.  
We view the former as more likely, because tidal barriers from a forming giant planet repel pebbles and planetesimals more efficiently than gas in protoplanetary disks (Tanaka \& Ida 1999, Paardekooper \& Mellema 2004). 
In that case, upward mixing is required to explain the fact that Jupiter, Saturn, Uranus and Neptune are all enriched in heavy elements compared to the Sun (e.g., Guillot \& Gautier 2014). 

This upward mixing of heavy elements, if convection is present, is energetically possible (Guillot et al. 2004). It requires these heavy elements to be miscible in the envelope, which appears to be the case (Wilson \& Militzer, 2010; 2012). 
An open question which remains is the efficiency at which this mixing (or core erosion) proceeds: this depends both on the initial state (formation mechanism), on the availability of overturning convection and on the efficiency of layered convection where it is present. 

Recently, the heavy-element distribution and core mass in proto-Jupiter at different stages during its formation was investigated (Lozovsky et al., 2017). 
The accreted planetesimals were followed as they entered the planetary envelope, and their distribution within the protoplanet accounting for settling (due to saturation) and convective mixing was determined. 
It was clearly shown that there is an important difference between the  Óheavy material massÓ $M_\mathrm{Z}$ and Ócore massÓ M$_{core}$, because most of the accreted heavy elements remain in the planetary envelope, and the core mass can be significantly smaller than the total heavy-element mass. 
This is demonstrated in Fig.~4 where we show M$_Z$ (red curve) vs.~ M$_{core}$ (purple curve). 
Although convective mixing can mix the heavy elements in the outer envelope, the innermost regions which have a steep enough composition gradient, can remain stable against convection.  
These inner regions can also consist of hydrogen and helium and could be viewed as
Ódiluted coresÓ. The Ódiluted coresÓ are of the order of 20 M$_{\oplus}$ in mass, but with lower density than that of a pure-Z core due to the existence of H+He. The left panel of Fig.~5 shows the calculated distribution of heavy-elements in proto-Jupiter based on formation models. 
\begin{figure}%
    \centering
    \subfloat[]{{\includegraphics[width=5.3cm]{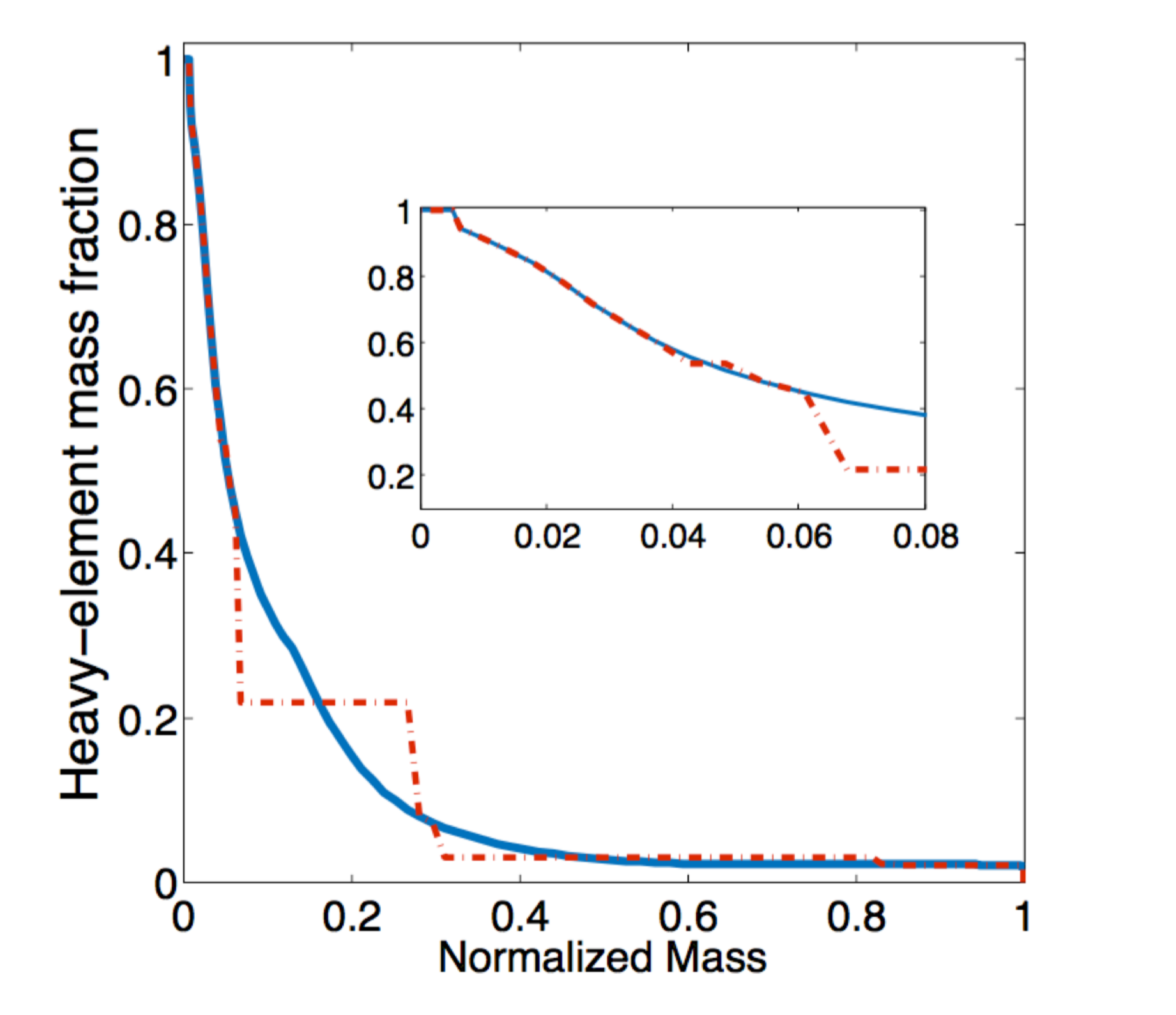} }}%
    \qquad
    \subfloat[]{{\includegraphics[width=4.95cm]{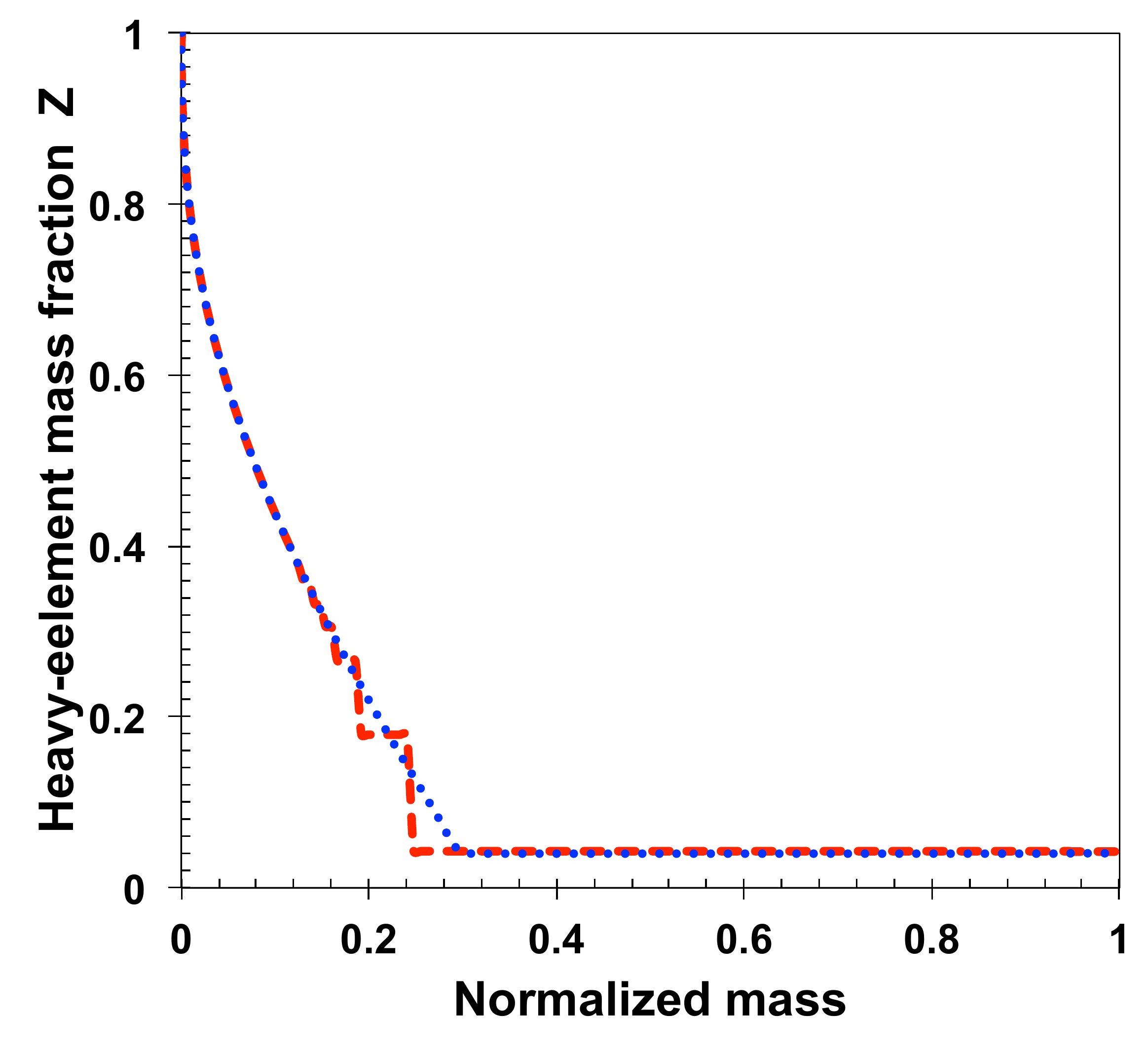} }}%
    \caption{
    {\small
    {\bf (a):} The distribution of heavy-elements in proto-Jupiter as it approaches its final mass at the end of rapid gas accretion.  The formation model corresponds to $\sigma$ = 10 g/cm$^2$ and 1 km-sized planetesimals (from Lozovsky et al., 2017). {\bf (b):} The primordial (dotted-blue) and current-state (dashed-red) distribution of heavy elements in Jupiter. The current-state model fits JupiterÕs measured $J_2$ and its estimated moment of inertia (see Vazan et al. 2016 for more details).
    }}%
    \label{fig:example}%
\end{figure}

\subsection{Mixing During Long-Term Evolution}
If the distribution of heavy elements is not homogenous due to the formation process, as suggested by Lozovsky et al.~(2017), it can affect the planetary thermal evolution. In addition, the primordial internal structure might change during the several 10$^9$ of evolution. 
Convective mixing of a gradual distribution of heavy elements in Jupiter's interior was investigated by Vazan et al. (2016). The primordial internal structure is somewhat similar to the one found by formation models. 
The right panel of Fig.~5 shows the preliminary (dotted blue) and final (after 4.5$\times$10$^9$ years, dashed red) heavy-element distributions in Jupiter accounting for mixing using a state-of-the-art planet evolution code (e.g., Vazan et al., 2016). In this model, Jupiter consists of $\sim$ 40 M$_{\oplus}$ of heavies. 
It is found that the innermost regions ($\sim$ 15\% of the mass) are stable against convection, and therefore, act as a bottleneck in terms of heat transport, while the outer envelope is convective throughout the entire evolution. 
The increasing temperature gradient between the innermost non-adiabatic and non-convective region and the outer convective region leads to a small penetration inward during the evolution, 
leading to a moderate heavy-element enrichment in the outer envelope as time progresses (see red curve in Fig.~4b). The innermost region which is highly enriched in heavy element has a lower entropy and is stable against (large-scale) convection during the entire evolution. 
As a result, the temperatures in the inner regions of the planet remain high while the outer ones can cool efficiently. 

\section{Exoplanets}

\subsection{Radii and bulk compositions}

Since the mid 90s, we know that gaseous planets exist in other planetary systems, which provides us with the opportunity to study giant planets more generally. 
Giant exoplanets are a complementary group to the Solar-System's outer planets - while the measurements are typically limited to mass and radius determination which provides their mean density - their large number provide us with  good statistics in terms of planetary bulk composition and the physical mechanisms governing the planetary evolution. 
\par

Most of the giant exoplanets with well-known masses and radii are ``hot Jupiters'', i.e., short period ($P< 10\,$days) Jupiter-mass planets ($0.5-10\,\rm M_{\rm Jup}$). A significant fraction of these objects $\sim 50\%$ are more inflated than predicted by standard evolution models of irradiated planets, which implies that another physical mechanism either slows the planets' cooling and contraction, or leads to an extra dissipation of energy in their interior (e.g., Guillot \& Showman, 2002; Laughlin et al., 2011). 

The bulk composition of exoplanets can be inferred by assuming a common mechanism inflating hot Jupiters (e.g., Guillot et al. 2006, Burrows et al. 2007) or by selecting only modestly irradiated giant exoplanets (Thorngren et al. 2016). The amount of heavy elements that they contain is inversely related to their size. Because giant planets are compressible, both the irradiation that they receive and their progressive contraction must be taken into account. 
Three main results can be extracted from these studies:
\begin{itemize}
\item[1.] While giant exoplanets above the mass of Saturn are generally mostly made of hydrogen and helium, some of them require surprisingly large masses of heavy elements (up to hundreds of M$_{\oplus}$, e.g., Moutou et al., 2013) or large ratios heavy elements to gas (as in the case of HD149026b which contains about $70\,M_\oplus$ of heavies for a total mass of $120\,M_\oplus$ -- see Ikoma et al. 2006). 
\item[2.] The ratio of the mass of heavy elements to the total planetary mass is negatively correlated with planetary mass (Thorngren et al. 2016), in agreement with formation of these planets by {\it core accretion}. 
\item[3.] The mass of heavy elements in hot Jupiters appears to be correlated with the metallicity of the parent star (e.g., Guillot et al., 2006, Burrows et al. 2007). However, this correlation is not statistically significant in the sample of weakly irradiated planets (Thorngren et al. 2016). 
\end{itemize}

Generally, it appears that at least some of the planets were able to efficiently collect the solids present in the disks. 
This is something that is not clearly explained by formation models, and is linked to the (unknown) efficiency of solid accretion during {\it Phase-3} and/or to late accretion of solids. Understanding this subject should improve significantly with PLATO thanks to the precise characterisation of a large number of transiting giant exoplanets, including those at large distances from their parent stars.

\subsection{Massive cores or enriched envelopes?}

Studies of the evolution of giant exoplanets typically assume the planets are made of a dense core and a solar-composition hydrogen-helium envelope. However, as discussed above, this is no more than a convenient simplification and this assumption is not justified. In fact, it is likely that, as for the giant planets in our Solar System, a significant fraction of the heavy elements are in the envelope rather than in a central core. Whether the heavies are mixed in the envelope or not has two effects: An enriched envelope has a larger molecular weight and shrinks more effectively than when heavy elements are embedded in a central core (e.g., Baraffe et al., 2008). This also means a larger opacity which, for hot Jupiters which are cooling through a thick radiative zone (Guillot et al., 1996), implies a less efficient cooling and contraction (Guillot, 2005; Vazan et al., 2013). When the enrichment is moderate (say, a few times solar) whether the heavy elements are embedded in the core or distributed in the envelope seems to have limited consequences. However, for larger enrichments, it can lead to an overestimate of the heavy-element mass required to fit the planetary radius (Baraffe et al. 2008, Vazan et al., 2013). 
\par

As for Jupiter and Saturn, the presence of potentially large amount of heavy elements can also lead to double-diffusive convection in the envelope. It has been proposed that this may account for the anomalously large size of some hot Jupiters (Chabrier \& Baraffe, 2007). This is unlikely however, for the same reasons as discussed previously, namely that overturning convection develops easily and should limit the extent of the double-diffusive region. In addition, the effect of the increased mass of heavy elements essentially compensates the effect of the delayed contraction on the planetary radius caused by compositional inhomogeneity (Kurokawa \& Inutsuka, 2015). 

The extent of whether the envelopes and atmospheres are significantly enriched or not is thus important both to better constrain the bulk compositions of the planets, but also to provide constraints to planet formation models. The ability to characterise giant planet atmospheres, and in particular determine their enrichment in heavy elements, would enable us to link interior and atmospheric compositions. This is crucial to understand the interior structure and formation of these planets. Currently, the possible determinations of chemical abundances are to be taken with extreme caution, both because of data quality and uncertainties on the presence of clouds (e.g., Deming \& Seager, 2017), but this situation should change in the near-future, in particular with JWST.

\section{Conclusions}
Characterising the planets in the outer Solar System is an ongoing challenge. 
Each planet has its special features and open research questions that are associated with its special nature. 
For Jupiter, we still try to get a better determination of its core mass and overall enrichment. Also for Saturn, we still need to better constrain its composition and structure, but with a focus on the role of helium rain and its cooling rate. 
The internal structures of Uranus and Neptune should be better determined, the source for the different structures and cooling rates of the planets still has to be resolved. In addition, understanding the connection between giant planet formation, evolution, and structure is still incomplete and is highly desirable.  
Ongoing and future space missions provide more constraints for structure models, and at the same time introduce new challenges and directions for exploration for modelers.  
\par

Several of the open questions have the potential to be solved in the fairly near future, in particular, in the following subjects:
{\bf (1)} Improvements in EoSs calculations and experiments. 
This will allow us to understand the behaviour of materials at high pressures and temperatures, and to discriminate among various EoSs.  
{\bf (2)} Significant improvements of the measurement of Jupiter's and Saturn's gravitational fields by the Juno (Bolton et al., 2017) and Cassini (Spilker, 2012) missions. 
With accurate measurements of the gravitational fields, and of the water abundance in the case of Jupiter, we will be able to reduce the parameter space of possible internal structures.  
{\bf (3)} The potential of sending a probe into Saturn's atmosphere and measuring the abundance of noble gases would allow us to understand enrichment mechanisms in giant planets, and their origins.
In the longer run, a mission dedicated for the ice giants (Uranus and/or Neptune) would bring new views of these icy planets.
{\bf (4)}  Additional and more accurate measurements of giant and intermediate-mass exoplanets. 
An overview of the variation in atmospheric composition of giant exoplanets and its connection to the host star's properties, and accurate determination of the planetary mean density will allow us to understand the nature of giant and icy planets in a boarder manner. 
\par

Clearly, we still have not solved all the mysteries related to gaseous planets, and much work is required. However, we expect new observations, exciting discoveries, and theoretical developments that will lead to a leap in understanding the origin, evolution, and interiors of this class of planetary objects.    

\section{References}
{\small
Alibert, Y., Mordasini, C., Benz, W.~\& Winisdoer, C.~(2005). Models of giant planet formation with migration and disc evolution. {\it A\&A}, 434, 343. \\
Baraffe, I., Chabrier, G., \& Barman, T.~(2008). Structure and evolution of super-Earth to super-Jupiter exoplanets: I. heavy element enrichment in the interior. {\it A\&A}, 482, 315.\\
Baraffe, I., Chabrier, G., Fortney, J.~\& Sotin, C.~(2014). Planetary Internal Structures. Protostars and Planets VI, Henrik Beuther, Ralf S. Klessen, Cornelis P. Dullemond, and Thomas Henning (eds.), University of Arizona Press, Tucson, 914 pp., 763.\\
Bolton et al.~(2017). Jupiter's interior and deep atmosphere: the first close polar pass with the Juno spacecraft. {\it Science}, submitted. \\ 
Burrows, A., Hubeny, I., Budaj, J.~\& Hubbard, W.~B.~(2007). Possible solutions to the radius anomalies of transiting giant planets. {\it ApJ}, 661, 502.\\
Chabrier, G.~\& Baraffe, I.~(2007). Heat transport in giant (exo)planets: A new perspective. {\it ApJL}, 661, L81. \\
Conrath, D. \& Gautier, D.~(2000). Saturn Helium Abundance: A Reanalysis of Voyager Measurements. {\it Icarus}, 144, 124. \\
Deming, D.~\& Seager, S.~(2017). Illusion and Reality in the Atmospheres of Exoplanets. {\it JGR Planets}, 122, 53.\\
Folkner, W.~M.~et al., 2017. Jupiter gravity field estimated from the first two Juno orbits. {\it GRL}, under review. \\
Fortney, J. J. \& Hubbard, W.~B.~(2003). Phase separation in giant planets: inhomogeneous evolution of Saturn. {\it Icarus}, 164, 228.\\
Fortney, J.~J.~\& Nettelmann, N.~(2010). The interior structure, composition, and evolution of giant planets. {\it Space Sci.~Rev.}~152, 423. \\
Fortney, J.~J., Helled, R., Nettelmann, N., Stevenson, D.~J., Marley, M.~S., Hubbard, W.~B.~\& Iess, L.~(2016). Invited review for the forthcoming volume "Saturn in the 21st Century." eprint arXiv:1609.06324 \\
Fuller, J.~(2014). Saturn ring seismology: Evidence for stable stratification in the deepinterior of Saturn, {\it Icarus}, 242, 283.\\
Guillot, T., Burrows, A., Hubbard, W.~B., Lunine, J.~I.~\& Saumon, D.~(1996). Giant planets at small orbital distances. {\it ApJL}, 459, L35.\\
Guillot, T.~(1999). A comparison of the interiors of Jupiter and Saturn. {\it Icarus}, 47. \\
Guillot, T.~(2005). The interiors of giant planets: Models and outstanding questions. {\it Annual Review of Earth and Planetary Sciences}, 33.\\
Guillot, T.~\& Showman, A.~P.~(2002) Evolution of "51 pegasus b-like" planets. {\it A\&A}, 385, 156.\\
Guillot, T., Santos, N.~C., Pont, F., Iro, N., Melo, C.~\& Ribas, I.~(2006). A correlation between the heavy element content of transiting extrasolar planets and the metallicity of their parent stars. {\it A\&A}, 453,  L21.\\
Guillot, T.~\& Gautier, D.~(2014).  Treatise on Geophysics (Eds. T. Spohn, G. Schubert). Treatise on Geophysics, 2nd edition.  \\
Helled, R.~\& Lunine, J.~(2014). Measuring jupiter's water abundance by juno: the link between interior and formation models. {\it MNRAS},  441, 2273.\\
Helled, R., Anderson, J.~D., Podolak, M.~\& Schubert G.~(2011). Interior models of Uranus and Neptune. {\it ApJ}, 726,15. \\
Helled, R., Anderson,  J.~D.~\& Schubert G.~(2010). Uranus and Neptune: Shape and rotation. {\it Icarus}, 210, 446. \\
Helled, R.~\& Guillot, T.~(2013). Interior models of Saturn: Including the uncertainties in shape and rotation. {\it ApJ}, 767, 113.\\
Hori, Y. \& Ikoma, M.~(2011). Gas giant formation with small cores triggered by envelope pollution by icy planetesimals. {\it MNRAS}, 416, 419. \\
Hubbard, W.~B.~\& Horedt, G.~P.~(1983). Computation of Jupiter interior models from gravitational inversion theory. {\it Icarus}, 54, 456.\\
Hubbard, W.~B. \& Militzer, B.~(2016). A preliminary Jupiter model. {\it ApJ}, 820, 80.\\
Iaroslavitz, E.~\& Podolak, M.~(2007). Atmospheric mass deposition by captured planetesimals. {\it Icarus}, 187, 600-.\\
Lambrechts, M.; Johansen, A.~(2014). Forming the cores of giant planets from the radial pebble flux in protoplanetary discs. {\it A\&A}, 572, id.A107, 12 pp.\\
Kurokawa, H. \& Inutsuka, S.~(2015). On the Radius Anomaly of Hot Jupiters: Reexamination of the Possibility and Impact of Layered Convection. {\it ApJ}, 815, 78.\\
Laughlin, G., Crismani, M.~\& Adams, F.~C.~(2011). On the anomalous radii of the transiting extrasolar planets. {\it ApJ}L, 729, L7.\\
Leconte, J.~\& Chabrier, G.~(2012). A new vision on giant planet interiors: the impact of double diffusive convection. {\it A\&A}, 540, A20\\
Leconte, J.~\& Chabrier, G.~(2013). Layered convection as the origin of SaturnÕs luminosity anomaly. {\it Nature Geoscience}, 6, 347.\\
Levison, H.~F., Kretke, K.~A. \& Duncan, M.~J.~(2016). Growing the gas-giant planets by the gradual accumulation of pebbles. {\it Nature}, 524, 322.\\
Lorenzen, W., Holst, B.~\& Redmer, R.~(2009). Demixing of Hydrogen and Helium at Megabar Pressures. {\it PRL}, 102(11), 115701.\\
Lorenzen, W., Holst, B.~\& Redmer, R.~(2011). Metallization in hydrogen-helium mixtures. {\it Phys.~Rev.~B}, 84(23), 235109.\\
Loubeyre,  P., Letoullec, R.~\& Pinceaux, J.~P.~(1991). A new determination of the binary phase diagram of H$_2$-He mixtures at 296 K. {\it Journal of Physics: Condensed Matter}, 3, 3183.\\
Lozovsky, M., Helled, R., Rosenberg, E.~D.~\& Bodenheimer, P.~(2017). JupiterÕs Formation and Its Primordial Internal Structure. {\it ApJ}, 836, article id. 227, 16 pp.\\
Mankovich, C., Fortney, J.~J.~\& Moore, K.~L.~(2016). Bayesian Evolution Models for Jupiter with Helium Rain and Double-diffusive Convection. {\it ApJ}, 832, article id. 113, 13 pp. \\
Marley,  M.~S., G{\'o}mez P.~ \& Podolak, M.~(1995). Monte Carlo interior models for Uranus and Neptune. {\it GJR}, 100,  23349.\\
Miguel, Y., Guillot, T.~\& Fayon, L.~(2016). Jupiter internal structure: the effect of different equations of state. {\it A\&A}, 596, id.A114, 12 pp. \\
Militzer, B., Hubbard, W.~B., Vorberger, J., Tamblyn, I.~\& Bonev, S.~A.~(2008). A massive core in Jupiter predicted from first-principles simulations. {ApJL}, 688, L45.\\
Militzer, B., Soubiran, F., Wahl, S.~M., Hubbard, W.~(2016). Understanding Jupiter's interior. {\it JGR: Planets}, 121, 1552-.\\
Mirouh, G. M., Garaud, P., Stellmach, S., Traxler, A. L., \& Wood, T. S.~(2012). {\it ApJ}, 750, 61.\\
Mizuno, H.~(1980). Formation of the giant planets. {\it Progress of Theoretical Physics}, 64, 544.\\
Morales, M. A., Hamel, S., Caspersen, K.~\&Schwegler,  E.~(2013). Hydrogen-helium demixing from first principles: From diamond anvil cells to planetary interiors. {\it Phys. Rev. B}, 87, 174105.\\
Morales, M. A.,Schwegler, E., Ceperley, D.~et al.~(2009). Phase separation in hydrogen-helium mixtures at Mbar pressures. {\it PNAS}, 106, 1324.\\
Nettelmann, N., Fortney, J.~J., Moore, K.~\& Mankovich, C.~(2014). An exploration of double diffusive convection in jupiter as a result of hydrogen-helium phase separation. {\it MNRAS}, 447, 3422.\\
Nettelmann, N., Helled, R., Fortney, J.~J., and Redmer, R.~(2012). New indication for a dichotomy in the interior structure of uranus and neptune from the application of modi ed shape and rotation data. {\it Planet.~Space Sci.}, special edition, 77, 143.\\
Nettelmann, N., Holst, B., Kietzmann, A., French, M., Redmer, R., and Blaschke, D.~(2008). Ab initio equation of state data for hydrogen, helium, and water and the internal structure of Jupiter. {\it ApJ}, 683, 1217.\\
Nettelmann, N., Becker, A., Holst, B.~\& Redmer, R.~(2012). Jupiter models with improved ab initio hydrogen equation of state (H-REOS.2). {\it ApJ}, 750, 52.\\
Nettelmann, N., P\"{u}stow, R.~\& Redmer, R.~(2013). Saturn layered structure and homogeneous evolution models with different EOSs. {\it Icarus} 225, 548.\\
Paardekooper, S.~J.~\& Mellema, G.~(2004). Planets opening dust gaps in gas disks. {\it A\&A}, 425, L9.\\
Podolak, M., Hubbard, W.~B.~\& Stevenson D.~J.~(1991). Model of UranusÕ interior and magnetic field. In: Uranus, 29Ð61. Tucson, AZ: University of Arizona Press\\
Podolak, M., Weizman, A.~\& Marley M.~S.~(1995). Comparative models of Uranus and Neptune. {\it PSS}, 43, 1517.\\
Podolak, M., Podolak J.~I.~\& Marley M.~S.~(2000). Further investigations of random models of Uranus and Neptune. {\it PSS}, 48, 143.\\
Podolak, M.~\& Helled, R.~(2012). What Do We Really Know about Uranus and Neptune? {\it ApJL}, 759, Issue 2, article id. L32, 7 pp.\\
Pollack, J.~B., Hubickyj, O., Bodenheimer, P., Lissauer, J.~J., Podolak, M.~\& Greenzweig, Y.~(1996). Formation of the Giant Planets by Concurrent Accretion of Solids and Gas. {\it Icarus}, 124, 62.\\
P\"{u}stow, R., Nettelmann, N., Lorenzen, W., \& Redmer, R.~(2016). H/He demixing and the cooling behavior of Saturn. {\it Icarus}, 267, 323.\\
Rosenblum, E., Garaud, P., Traxler, A.~\& Stellmach, S.~(2011). Erratum: "Turbulent Mixing and Layer Formation in Double-diffusive Convection: Three-dimensional Numerical Simulations and Theory". {\it ApJ}, 742, 132\\
Saumon, D.~\& Guillot, T.~(2004). Shock compression of deuterium and the interiors of Jupiter and Saturn. {\it ApJ}, 609, 1170.\\
Schouten,  J.~A.,  de Kuijper,  A.~\& Michels,  J.~P.~J.~(1991). Critical line of He-H$_2$ up to 2500 K and the influence of attraction on fluid-fluid separation.  {\it Phys. Rev. B}, 44, 6630.\\
Spilker, L.~J.~(2012). Cassini: Science highlights from the equinox and solstice missions. In: Lunar and Planetary Institute Science Conference Abstracts, 43, p. 1358.\\
Stevenson, D.~J.~\& Salpeter E.~E.~(1977a). The dynamics and helium distribution in hydrogen-helium fluid planets. {\it ApJS}, 35, 239.\\
Stevenson, D.~J.~\& Salpeter, E.~E.~(1977b). The phase diagram and transport properties for hydrogen-helium fluid planets. {\it ApJS}, 35, 221.\\
Tanaka, H. \& Ida, S.~(1999). Growth of a Migrating Protoplanet. {\it Icarus}, 139, 350\\
Thorngren, D.~P., Fortney, J.~J., Murray-Clay, R.~A.~\& Lopez, E.~D.~(2016). The Mass-Metallicity Relation for Giant Planets. {\it ApJ}, 831, article id. 64, 14 pp.\\
Vazan, A., Helled, R., Kovetz, A.~\& Podolak, M.~(2015). Convection and Mixing in Giant Planet Evolution. {\it ApJ}, 803, 32.\\
Vazan, A., Helled, R., Podolak, M.~\& Kovetz, A.~(2016).  The Evolution and Internal Structure of Jupiter and Saturn with Compositional Gradients. {\it ApJ}, 829, 118.\\
Venturini, J., Alibert, Y., \& Benz, W.~(2016). {\it A\&A}, 596, id.A90, 14 pp.\\
von Zahn, U., Hunten, D.~M.~\& Lehmacher, G..~(1998). Helium in JupiterÕs atmosphere: Results from the Galileo probe helium interferometer experiment. {\it JGR}, 103, 22815.\\
Wahl, Sean M et al..~(2017). {\it GRL}, submitted. \\
Wilson, H.~F.~\& Militzer, B.~(2010). Sequestration of noble gases in giant planet interiors. {\it PRL}, 104,  121101.\\
Wilson, H.~F.~\& Militzer, B.~(2012). Solubility of water ice in metallic hydrogen: Consequences for core erosion in gas giant planets. {\it ApJ}, 745, 54.\\
Wood, T. S., Garaud, P. \& Stellmach, S.~(2013). A new model for mixing by double-diffusive convection (semi-convection). II. The transport of heat and composition through layers. {\it ApJ}, 768, 157.\\
}
\end{document}